\documentclass[%
 aip,
 amsmath,amssymb,
 reprint,%
]{revtex4-1}
\usepackage{float}
\usepackage{graphicx}% Include figure files
\usepackage{dcolumn}% Align table columns on decimal point
\usepackage{bm}% bold math
\usepackage[utf8]{inputenc}
\usepackage[T1]{fontenc}
\usepackage{mathptmx}

\begin{document}

\preprint{AIP/123-QED}

\title{A 2D scintillator based proton detector for HRR experiments}
% Force line breaks with \\

\author{M. Huault}

\affiliation{ 
	Centro de Laseres Pulsados, Building M5, Science Park, Calle Adaja 8, 37185 Villamayor, Salamanca, Spain%\\This line break forced with \textbackslash\textbackslash
}%
\affiliation{Universidad de Salamanca, Patio de Escuelas 1,  37008 Salamanca, Spain}%

\author{D. De Luis}
\author{J. Alpinaniz}
\author{M. De Marco}
\author{J A. P{\'e}rez-Hern{\'a}ndez}
\affiliation{ 
	Centro de Laseres Pulsados, Building M5, Science Park, Calle Adaja 8, 37185 Villamayor, Salamanca, Spain%\\This line break forced with \textbackslash\textbackslash
}%
\author{N. Gordillo Garc\'{i}a}
\author{C. Guti\'{e}rrez}
\affiliation{CMAM, Universidad Aut\'{o}noma Madrid, Campus de Cantoblanco, E-28049 Madrid, Spain%Second institution and/or address%\\This line break forced% with \\
}%
\author{L. Roso}
\affiliation{ 
	Centro de Laseres Pulsados, Building M5, Science Park, Calle Adaja 8, 37185 Villamayor, Salamanca, Spain%\\This line break forced with \textbackslash\textbackslash
}%
\affiliation{Universidad de Salamanca, Patio de Escuelas 1,  37008 Salamanca, Spain}%
\date{\today}% It is always \today, today,
\author{G. Gatti}
\affiliation{ 
	Centro de Laseres Pulsados, Building M5, Science Park, Calle Adaja 8, 37185 Villamayor, Salamanca, Spain%\\This line break forced with \textbackslash\textbackslash
}%
\author{L. Volpe}

\affiliation{ Centro de Laseres Pulsados, Building M5, Science Park, Calle Adaja 8, 37185 Villamayor, Salamanca, Spain%\\This line break forced with \textbackslash\textbackslash
}%
\affiliation{Laser-Plasma Chair at the University of Salamanca, patio de escuelas 1, Salamanca, Spain}%
\email{lvolpe@clpu.es}
\begin{abstract}
We present a scintillator based detector able to measure both spatial and energy information at High repetition rate (HRR) with a relatively simple design. It has been built at the Center of Pulsed Laser (CLPU) in Salamanca and tested in the proton accelerator at the Centro de Micro-Análisis de Materiales (CMAM) in Madrid. The detector has been demonstrated to work in HRR mode by reproducing the performance of the radiochromic film detector. It represents a new class of on-line detectors for Laser-plasma physics experiments in the new emerging High Power and HRR laser systems.

\end{abstract}

\maketitle

\section{State of the art}

The advent of High power lasers working at High repetition rate is nowadays a reality and HRR proton sources are now routinely produced with energies ranging from few to tens of MeV. Laser-driven proton sources are characterised by a divergence that in several measurements has been proved to be related with the energy of the protons and the spatial distribution of the proton beam \cite{1Snavely2000}. 

\begin{figure} [h]
	\includegraphics[scale=0.31]{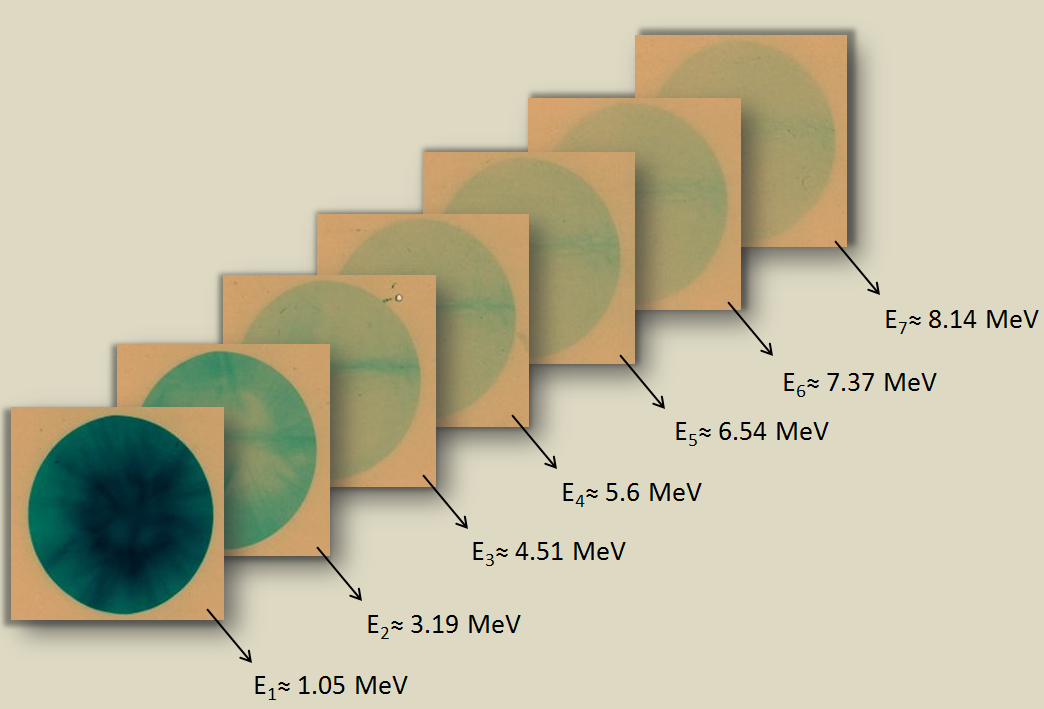}
	\caption{Sample of a Gafchromic HD-V2 radiocrhomic films irradiated at Center of Laser pulsed CLPU facility (Spain) by a proton beam.}
\label{fig1} 
\end{figure}

Laser-driven proton beam are becoming more and more important for several applications in different fields of physics \cite{2Fritzler2003}, chemistry and material science \cite{3surface,4Johansson1986} as well as biology, medicine \cite{5Malka2004} and cultural heritage \cite{6Barberio2017}. For this the spatial and energy characterisation of the proton beams is becoming more and more important for the potential use of such sources. The first demonstration of laser-driven protons production was carried out  in laser system working at single shot mode and one of the most used diagnostics consist in a series of  Radiochromic films \cite{7Green2016} placed one after each other and able to recover spatial distribution as a function of the energy (see figure \ref{fig1}). 

The possibility to extend this technique to HRR mode of operation is nowadays a challenge in laser plasma community and several laboratories and research group are working on this . The main idea is to substitute the active RCF layers  with scintillator detectors capable to transform ion energy deposition in light that can be then  collected by an optical ccd camera. Several research group have proposed special on-line configurations to imitate the RFC stack but up to now only a partial extension of the RCF capabilities was possible. During 2011 and 2012 two research groups from United Kingdom and from Germany have proposed scintillator based detectors. The group from Rutherford Appleton Laboratory (RAL) \cite{8Green2011} proposed to use detectors sensitive to different wavelength which is limited only to three wavelengths and it is extremely complicated in the mode of working of the acquisition system. The group from Dresden \cite{9Metzkes2012,10Metzkes2016} proposed a stack of scintillators placed one after each other as the RCFs stack with a system to read the scintillation alonge the transversal direction; this detector have been tested in an proton beam accelerator and currently is used in the Dresden laboratory. Both the detectors can reproduce partially the mode of working of the RCFs stacks even if they are increasing the complexity of the viewing system and the data interpretation.

\section{Explanation of the device}
Here we present a scintillator based detector able to measure both spatial and energy information at HRR with a relatively simple design. It consist in a series of scintillators placed similarly as an RCF stack (as shows in fig.\ref{fig1}) but with a relative angle one respect the others to let the necessary field of view for an imaging system looking at the back side of each layer. The Imaging system can be arranged depending on the special condition and is not a critical part of the device. Each of the scintillator plate is covered by an aluminium foil to protect it by the light emission from the previous scintillator plate. 
The relative angle between each layer is the key factor in the design because it permit the acquisition of the full 2D proton distribution for each of the layer composing the stack. It is a relevant parameter because the total size of the detector depends critically on it. Indeed increasing such angle the total length of the detector increase and then the final proton emission solid angle also increase with the scintillator area. The optimal design of the detector is a trade off between $\phi$ that is the relative half angle between 2 consecutive plates and the dimension of the scintillator stack compared to the maximum accepted energy. 

\begin{figure*}[ht]
	\includegraphics[scale=0.5]{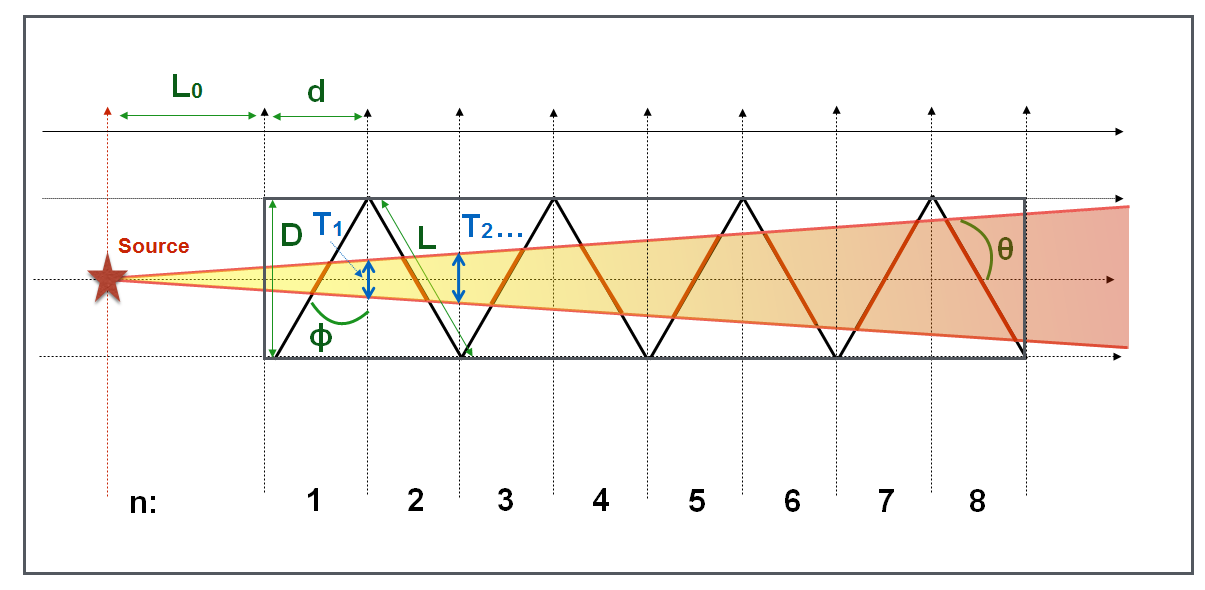}
	\caption{2D top view of detector; the proton beam solid angle is parametrised by the internal half angle $\theta$, the detector dimension D is represented by the length of the scintillator plate L and the relative half angle btween the plates $\phi$, n is the number of layers, $L_0$ is the distance between the proton sources and the detector, d the longitudinal dimension of the scintillator plate and $T_1$...$T_n$ represent the projection of the proton beam solid angle for each plate. }
\label{fig2} 
\end{figure*}

More in detail let assume a proton beam propagating in a symmetric cone with half angle $\theta$ (see figure \ref{fig2}), where the transversal and longitudinal dimension of each scintillator plate (which here are assumed egual) perpendicular to the proton beam direction can be written as: 
\begin{eqnarray}
D=L \text{cos}(\phi)
\label{eq1}
\end{eqnarray} 
 and \begin{eqnarray}
 d=L \text{sin}(\phi)
 \label{eq2}
 \end{eqnarray}
The projection of proton cone in the scintillator plate can be described as: 
\begin{eqnarray}
T_n(\theta,\phi)= 2L_n(\phi) \text{tan}(\theta)
\label{eq3}
\end{eqnarray}

\begin{eqnarray}
\text  where \quad  L_{n}(\phi)=L_{0}+nL\sin(\phi)
 \label{eq4}
 \end{eqnarray}

$L_n$ is the effective length of the detector considered from the proton source emission. Lets note that $L_0=0$ is less realistic case because a minimum distance between detector and source must be accounted for to host a magnet to deflect electrons which are generated in the process to do not affect the scintillation signal. Finally $L_n$  depends both i) on the angle between two successive plates $\phi$, ii)  on the dimension $L$ and iii) on the number $n$ of the scintillator foils. 

The working condition can be written as $D> T_n(\theta,\phi)$ i.e the size of transverse projection of the scintillator D must be larger than the projection of proton solid angle $T_n$. This can be solved as:

\begin{eqnarray}
&&n < n_0(\theta,\phi)+n'(L_0,L,\phi)\\
&& n_0(\theta,\phi)=\frac{1}{2\text{tan}\left[ \theta \text{tan}\phi \right]}\nonumber \\
&&n'(L_0,L,\phi)=-\frac{L_0}{L}\frac{1}{\text{sin}\phi} \nonumber
\label{eq5}
\end{eqnarray}

Where $n_0=n(\theta,\phi,L_0=0,L)$

\subsection{case $L_0=0$}
The case $L_0=0$ corresponds to assume the proton source just placed in the surface corresponding to n=0 so the system becomes: 

\begin{eqnarray}
n < n_0(\theta,\phi)
\label{eq6}
\end{eqnarray}

Equation \ref{eq6} can be studied as a function of  $\phi$  (for a given value of $\theta$ here 25, 20, 15 and 10  degrees).

\begin{figure}[hbt!]
	\includegraphics[scale=0.87]{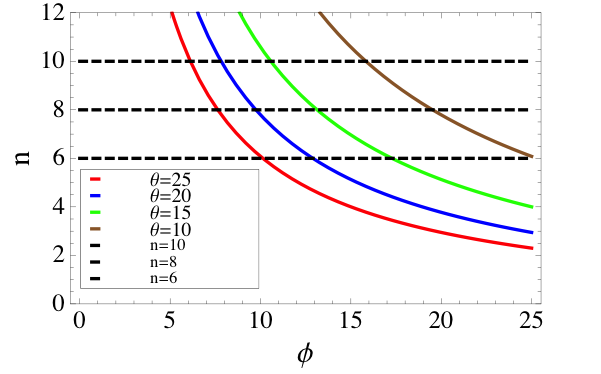}
	\caption{Eq.\ref{eq6} is represented in the figure. The number of layers n as a function of the half angle between the plates $\phi$ for different divergence half angles $\theta$ is plotted and compared for different value of n. When the curves n($\phi$) are above a given fixed n value (doted lines) the design of the detector is such that proton energies corresponding to the n value are detectable. As example for a proton beam with a 40 degrees of divergence ($\theta$ = 20º), 6 layers can work with a maximum angle $\phi \sim 13^{\circ} $,  and for 8 layers $\phi \sim 10^{\circ} $. It is important to note that the proton energy corresponding to the nth layer depends on the thickness and composition of the layer.}
	\label{fig3} 
\end{figure}

Figure \ref{fig3} shows the number of layers $n$ (representing eq. \ref{eq6}) as a function of $\phi$ for different divergence half angles $\theta$. The relative angle between the layers is a key parameter for the detector design and it value needs to be reduced as much as possible to maximise the possible number of layers maintaining a reasonable dimension of the detector. In addition $\phi$ cannot be below $10^\circ$ to let the imaging system work. 

\subsection{case $L_0 \ne 0$}
The most realistic case is $L_0 \ne 0$, lets assume proton divergence with a maximum half angle of 20$^{\circ}$ and according to fig. \ref{fig3}, let assume  a relative half angle between 2 plates $\phi=15^{\circ}$. Considering the above mentioned parameters we can  represent the eq. \ref{eq6}  as a function of $L_n$ for different values of $L_0$.  

\begin{figure}[hbt!]
	\includegraphics[scale=0.56]{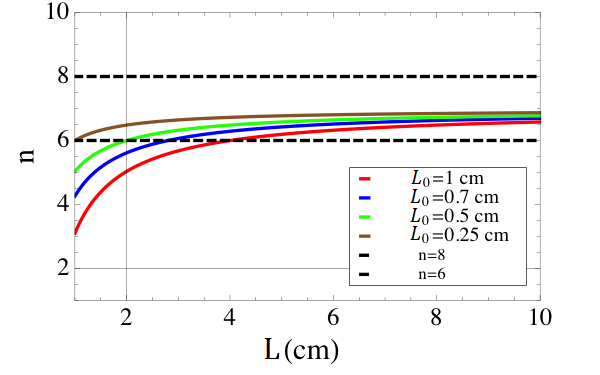}
	\caption{The number of scintillator layers $n$ is represented as a function of the scintillator foil  length $L$ for different values of $L_0$}
\label{fig4} 
\end{figure}

\begin{figure}[hbt!]
	\includegraphics[scale=0.56]{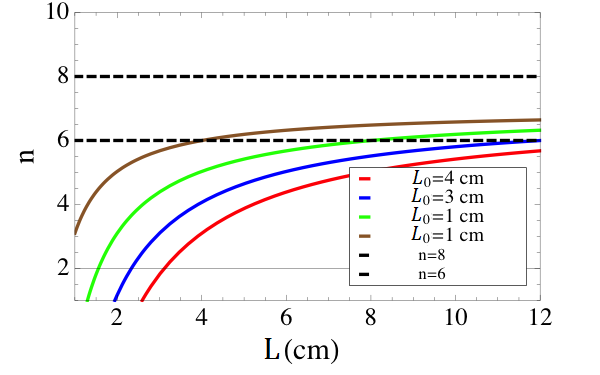}
	\caption{similarly to fig.\ref{fig3} the number of scintillator layers $n$ is represented as a function of the scintillator foil  length $L$ for larger values of $L_0$}
\label{fig5} 
\end{figure}

The final result shows that the interplay between the relative angle $\phi$ and the initial  distance $L_0$ must be carefully adjusted to optimise the design. This limit the range of possible proton energies  to be detected from the detector in one single shot but we will see that such limitations can be overcome with special and dedicated adjustments of the internal structures. In addition most of the experimental data shows that proton divergence reduce by increasing the proton energy and this will aslo mitigate the final constrains. On the contrary the detector size L is not playing a critical role  for the design and can be fixed around 20-30 ,mm, this fact is very relevant in maintaining the size of the detector reasonable for the dimension of the interaction chamber . Fig. \ref{fig5} shows a similar case with larger values of $L_0$, the result is that by increasing $L_0$ increase proportionally $L$ and of course $L_n$. 

\section{Test of the device}
A first detector prototype has been constructed at the Centro de Laseres Pulsados (CLPU) in Salamanca and tested at the Centro de Micro-Análisis de Materiales (CMAM) of the Universitad Autonoma de Madrid where a collimated proton beam up to 10 MeV is available for user access.

\begin{figure}[hbt!]
	\includegraphics[scale=0.56]{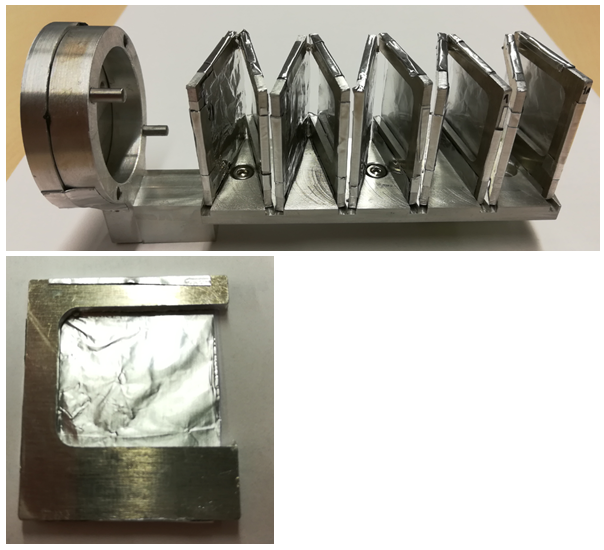}
	\caption{ Top image: Lateral view of the detector with a longitudinal dimension of the base between the first and the last plate = 90 mm ($L_{10}$ = 90 mm with L$_{0}$ = 0). Each plates are separated from each other with a relative angle $\phi$ = 12.5 º;  Bottom image: front view of the holder and its scintillator plate with a dimension of L = 20 mm.} 
	\label{fig6} 
\end{figure}

Fig.\ref{fig6} shows a customised version of the detector for using it in the accelerator in madrid with all the relevant parameters, the proton beam was 10 MeV energy with a $\Delta$E/E < 1 $\%$. The detector is made by 10 plates of scintillators BC-400 placed one after each other with an angle of 25º between them ($\phi$=12.5º). Each plate is about $150 \ \mu$m $\pm 50\  \mu$m thick with a free detector area of 20 mm $\times$ 20 mm. With this configuration, the detector can resolve energy larger than the maximum 10 Mev achievable in CMAM.

\begin{figure}[hbt!]
	\includegraphics[scale=0.55]{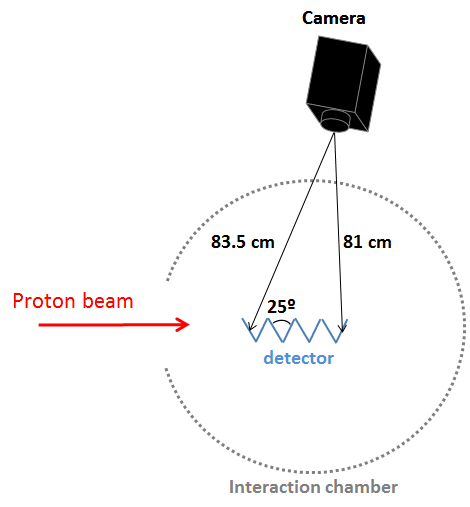}
	\caption{ Top view of the interaction chamber with the detector place inside and the camera set outside the chamber for recording the signal.   } 
	\label{fig7} 
\end{figure}

The detector was placed in the middle of the interaction chamber, on the front part of a 4-axis goniometer, able to rotate 360º around the propagation axis of the proton beam. The emission was collected from the back side of each scintillator
with a CCD camera Point Grey Blackfly monochrome model and an objective NIKON AF-S DX NIKKOR 18-105 mm f/3.5-5.6G ED VR placed outside the chamber at about 83.5 cm $\pm$ 0.5 cm from the first plate and 81 cm $\pm$ 0.5 cm from the tenth plate (see fig.\ref{fig7} ). 

\begin{figure}[hbt!]
	\includegraphics[scale=0.4]{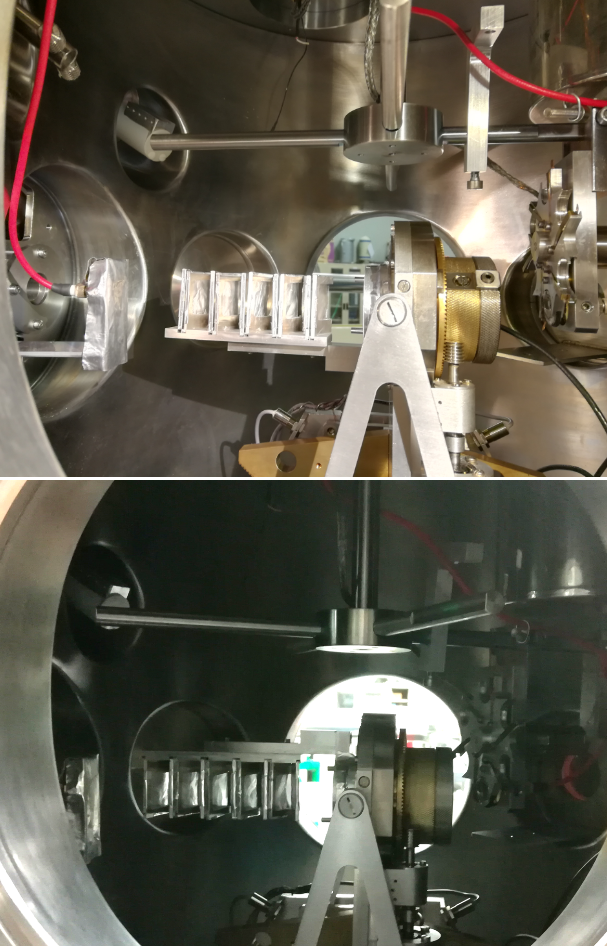}
	\caption{ Top picture represents the configuration 1, where the odd number scintillator plates are imaged by the camera. Bottom picture represents the configuration 2 with the imaging of the peer scintillator plates.  } 
	\label{fig8} 
\end{figure}

\begin{figure*}[hbt!]
	\includegraphics[scale=0.35]{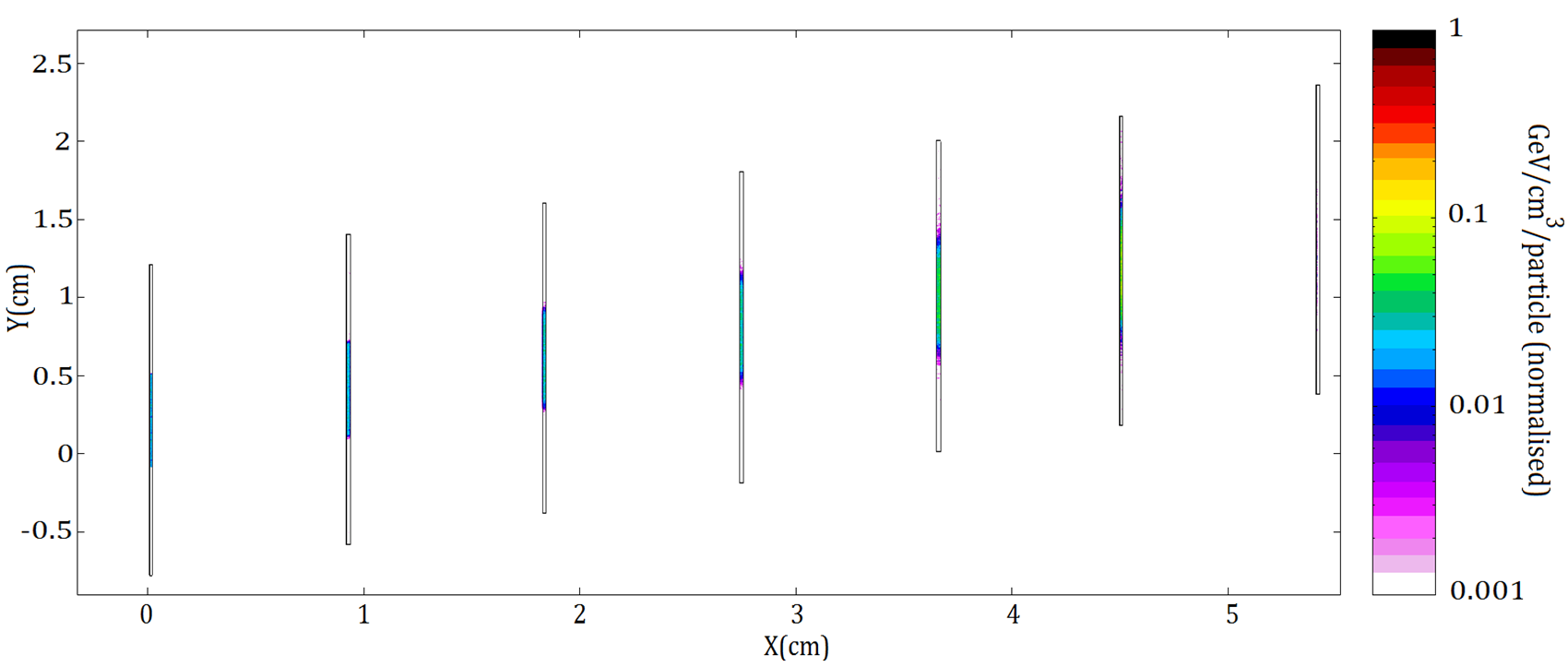}
	\caption{ Simulation of the deposited energy with FLUKA. the proton beam arrive with an incident angle of 15º on each plate. The scintillator plates are parallel to each other for simplicity of simulation but does not affect the resulTS. The colorscale defines the energy lost by 10 MeV proton beam in each scintillator. The axis X and Y represent the spatial distribution of the deposited energy (not at the scale for easier visualisation).} 
	\label{fig9} 
\end{figure*}

Since the proton source was considered very stable, 2 configurations of irradiation were done to be able to image the full detector with the same camera. The odd plates with the numbers 1,3,5,7 and 9 were pictured when the goniometer was in normal position (rotation axe at 0º) and the peer plates numbers 2,4,6,8 and 10 when the goniometer was at 180º of rotation (see figure \ref{fig8}). 10 MeV proton beam have been irradiating the detector under $10^{-6}$ mbar vacuum. Figure \ref{fig9} is a simulation on the deposited energy by a beam of 10 MeV proton on FLUKA in order to reproduce the experimental results. It is expected to obtain the bragg peak withing the 6th plate and the aluminium filter of the 7th plate.

\begin{figure*}[hbt!]
	\includegraphics[scale=0.59]{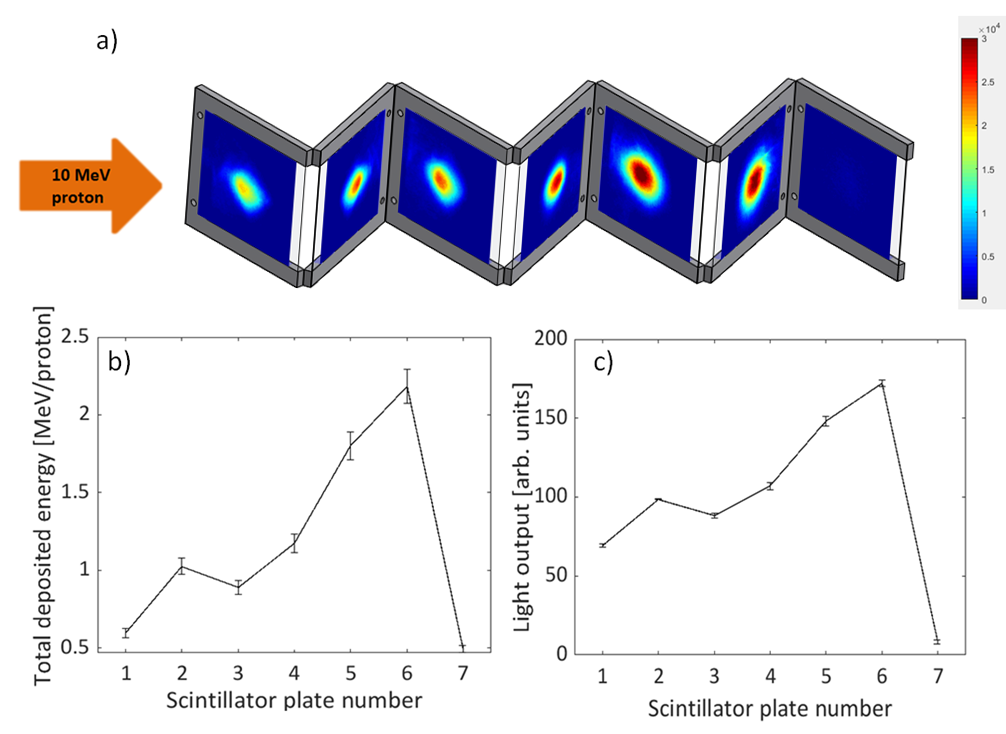}
	\caption{Picture a) re-constructed signal obtained by the CCD camera during the irradiation with a 10 MeV proton beam with the color scale in pixel value. Figure b) is obtained with FLUKA simulation (see fig.9) and represents the total deposited energy per particle for each scintillator plate irradiated by a 10 MeV proton beam. Figure c) represents the response of the scintillator (light output) to 10 MeV proton beam irradiation.} 
	\label{fig10} 
\end{figure*}
Figure \ref{fig10}a shows the signal recorded during the two configurations of irradiation and have been re-built to get the signal in a single image. The total deposited energy per proton  for each scintillator plate has been extracted from FLUKA simulation and is represented in figure \ref{fig10}b. It is in good agreement with the scintillator response in figure \ref{fig10}c). The light output has been obtained from recorded pictures by the conversion of each pixel value into photon. We can observed a slight flattening of the  scintillator response around the bragg peak (high stopping power) that can be interpreted as a saturation of the response  and is due to the quenching effect. A paper will be published about its calibration in detail very soon where it has been confirmed a linear response for proton flux but non-linearity response to the absorbed dose for high stopping power proton energy. Knowing the corrective factor of this effect \cite{11Torrisi2000}, the detector can be suitable for 2D spatial distribution measurements.

\section{\label{sec:level1}Discussion and conclusions}

A scintillator based 2D ion detector for high repetition rate experiments has been designed, builted  at the CLPU and tested in a first ideal condition in the proton accelerator at the CMAM in Madrid.
The detector has been demonstrated to work in HRR mode by fully reproducing the performance of the consolidated RCF detector. With a detailed analysis we have shown that it is possible to account for the laser-driven proton divergence by maintaining a compact size of the detector and that is possible to remove electron signal by placing a relatively small size magnet in front of the detector entrance.
Finally the presented design of 2D ion detector is promising for substitute the classical RCF stack detector at HRR mode of working. It represent a new class of on-line detectors to developed Laser-plasma physics experiments in the new emerging High Power and HRR laser systems.

\begin{acknowledgments}
Authors acknowledge to FURIAM project FIS2013-4774-R, PALMA project FIS2016-81056-R, LaserLab Europe IV Grant No. 654148, from Junta de Castilla y Le{\'o}n Grant No. CLP087U16 and the Unidad de Investigaci{\'o}n Consolidada (UIC) 167 from Junta de Castilla y Le{\'o}n.
\end{acknowledgments}

\end{document}